# Enhancement of the energy storage and electrocaloric effect performances in 0.4 BCZT–0.6 BSTSn medium-entropy ceramic prepared by sol-gel method.

S. Khardazi[1*], Z. Gargar[2], A. Lyubchyk[3], O. Zakir[4], D. Mezzane[1,3], M. Amjoud[1], A. Alimoussa[1], Z. Kutnjak[5]

[1] IMED-Lab, Cadi-Ayyad University, Faculty of Sciences and Technology, Department of Applied Physics, Marrakech, 40000, Morocco

[2] Laboratoire des Sciences des Matériaux et Optimisation des Procédés, Faculté des Sciences Semlalia, Université Cadi Ayyad, Marrakech, Morocco

[3] DeepTechLab, Faculty of Engineering, Lusofona University, 376 Campo Grande, 1749-024, Lisbon, Portugal

[4] Physical-Chemistry of Materials and Environment Laboratory, Faculty of Sciences Semlalia, Cadi Ayyad University, Marrakech, Morocco

[5]Jožef Stefan Institute, Ljubljana, 1000, Slovenia

**Abstract**

Based on the traditional polycrystalline ferroelectric $Ba_{0.85}Ca_{0.15}Zr_{0.10}Ti_{0.90}O_3$, the 0.4 $Ba_{0.85}Ca_{0.15}Zr_{0.10}Ti_{0.90}O_3$ –0.6 $Ba_{0.9}Sr_{0.1}Ti_{0.9}Sn_{0.1}O_3$ medium-entropy material with good energy storage and electrocaloric effect performances is designed and synthesized by sol-gel method. The structural, dielectric, energy storage and electrocaloric effect properties of the prepared sample were studied. The findings demonstrate that the 0.4 $Ba_{0.85}Ca_{0.15}Zr_{0.10}Ti_{0.90}O_3$ –0.6 $Ba_{0.9}Sr_{0.1}Ti_{0.9}Sn_{0.1}O_3$ ceramic simultaneously has a significant recoverable energy storage density of 255.4 mJ/cm$^3$, an efficiency of 67.9%, a large electrocaloric effect temperature change of $\Delta T$ = 1.36 K, and a high $\xi_{max}$ of 0.453 K·mm/kV under a low electric field of 30 kV/cm. Moreover, excellent temperature stability (40–120 °C) of the recoverable energy storage $W_{rec}$ (less than 10%) was achieved in the investigated sample 0.4BCZT-0.6BSTSn. This study demonstrates that the 0.4BCZT-0.6BSTSn ceramic is a promising candidate for solid-state cooling and energy storage dielectric ceramics through exploring medium-entropy composition.

* Corresponding author.

E-mail address: khardazisaid@gmail.com, said.khardazi@ced.uca.ma (S. Khardazi).

Tel: +212 608219671

## 1. Introduction

Energy storage technologies are vital for the efficient use of renewable energy sources and the stabilization of electrical grids. Among various energy storage materials, dielectric ceramics have garnered significant attention due to their excellent energy storage density [1- 3]. The dielectric capacitor appears to be well-positioned to meet some of tomorrow's energy needs, such as high power systems, pulse applications, electronic devices, etc., because of its fast charging and discharging rate (~ μs scale), long cycle life (>$10^6$), and good reliability, and the growing need for new energy storage modalities [4- 7]. Traditionally, lead-based ceramics, such as lead zirconate titanate (PZT), have been widely studied for energy storage applications due to their superior dielectric properties. However, the environmental and health hazards associated with lead have driven the scientific community to explore lead-free alternatives.

Recently, high- and medium-entropy engineering has been considered as one of the interesting methods to improve the energy storage capabilities of ceramics made of five or more ions [8- 10]. This method is based on entropy-driven phase stability and lattice distortion carried due to the high degree of atomic disorder, which results in improved overall characteristics [11]. Furthermore, the $ABO_3$ perovskite structure exhibits lattice distortion due to the enhancement of disorder and oxygen octahedral complexity inside the system lattice by the insertion of multiple ions with distinct valence states and radii at the same site [12]. The lattice deformation can be used to enhance the energy storage performance by tuning its structural characteristics [13]. For instance, several ceramics made of high-entropy perovskite oxide have been investigated, including $(Bi_{0.2}Na_{0.2}K_{0.2}Ba_{0.2}Ca_{0.2})TiO_3$ [3], $(Na_{0.2}Bi_{0.2}Ca_{0.2}Sr_{0.2}Ba_{0.2})TiO_3$ [14] and the composite (1-x) $(0.94Bi_{0.5}Na_{0.5}TiO_3$-$0.06BaTiO_3)$-x$(0.96NaNbO_3$-$0.04CaSnO_3)$ [15]. The development of lead-free ceramics is crucial to creating environmentally friendly and sustainable energy storage systems. Over 70 years after its discovery, the market-dominant material $BaTiO_3$ (BTO) is the most extensively researched ferroelectric (FE) material. In addition, both the commercial and academic markets are highly interested in it (more than 3 trillion ceramic capacitors are produced annually using BTO-based materials) [1]. Moreover, to the best of our knowledge, only a few studies have reported on the energy storage and electrocaloric properties of $BaTiO_3$-based high-entropy ceramics [16]. Based on this purpose, the most sophisticated technique used to achieve high energy storage efficiency and large electrocaloric effect in lead-free ceramics consists of disrupting the long-range order of A-O and/or B-O coupling by introducing chemical inhomogeneity [17]. Some examples of this include $BaTiO_3$-$Bi(Mg_{1/2}Zr_{1/2})O_3$ [17], $0.87BaTiO_3$- $0.13Bi(Zn_{2/3}(Nb_{0.85}Ta_{0.15})_{1/3})O_3$ [18] and $0.7BaTiO_3$-$0.3BiScO_3$ [19]. It has been reported that the BT-based materials, $Ba_{0.85}Ca_{0.15}Zr_{0.10}Ti_{0.90}O_3$ (BCZT), exhibits a high piezoelectric coefficient ($d33\sim620$ pCN$^{-1}$), which is even greater than that of the PZT system ($d_{33}$ = 500–600 pC/N) due to the presence of the morphotropic phase boundary (MBP) region [20, 21]. Thus, the development of BCZT relaxor ceramics has prompted a greater interest in energy storage capacities and electrocaloric effects [22-25]. Based on our previous studies on the (1-x) $Ba_{0.85}Ca_{0.15}Zr_{0.10}Ti_{0.90}O_3$– $xBaTi_{0.89}Sn_{0.11}O_3$ ceramics synthesized by solid state [23] and sol-gel [24] methods, the composition x = 0.6 exhibited enhanced ferroelectric and energy storage properties. Meanwhile, it has been reported recently that the same composition (x = 0.6 ) showed an improved electrocaloric effect performance [27].

A medium-entropy approach was used in our investigation to optimize the energy storage capabilities and electrocaloric effect of $0.4Ba_{0.85}Ca_{0.15}Zr_{0.10}Ti_{0.90}O_3$–$0.6Ba_{0.9}Sr_{0.1}Ti_{0.9}Sn_{0.1}O_3$ ceramic by mixing $Ba_{0.9}Sr_{0.1}Ti_{0.9}Sn_{0.1}O_3$ (BSTSn) and $Ba_{0.85}Ca_{0.15}Zr_{0.10}Ti_{0.90}O_3$ (BCZT) exhibiting three kinds of elements occupying the A-site and the same number at the B-site to break the long-range ferroelectric to increase the configuration entropy and generate more polar nanoregions (PNRs).

## 2. Experimental Procedure

The $Ba_{0.85}Ca_{0.15}Zr_{0.10}Ti_{0.90}O_3$ (BCZT), $Ba_{0.9}Sr_{0.1}Ti_{0.9}Sn_{0.1}O_3$ (BSTSn) and 0.4 $Ba_{0.85}Ca_{0.15}Zr_{0.10}Ti_{0.90}O_3$ –0.6 $Ba_{0.9}Sr_{0.1}Ti_{0.9}Sn_{0.1}O_3$ (abbreviated as 0.4BCZT–0.6BSTSn) were successfully designed and synthesized by the sol-gel technique. BCZT and BSTSn were prepared individually and then mixed in a suitable weight ratio to acquire the desired composite ceramic 0.4BCZT–0.6BSTSn. The chemical reagents barium acetate $Ba(CH_3COO)_2$, strontium acetate $Sr(CH_3COO)_2$, tin chloride dihydrate $SnCl_2\cdot2H_2O$, titanium isopropoxide ($C_{12}H_{28}O_4Ti$), zirconium oxychloride ($ZrOCl_2\cdot8H_2O$) and calcium nitrate tetrahydrate ($Ca(NO_3)_2\cdot4H_2O$) were an analytical grade and used as initial precursors without further purification. Acetic acid $CH_3CH_2OOH$ and 2-methoxyethanol ($C_3H_8O_2$) were used as solvent agents. The details of the BCZT and BSTSn powder synthesis process can be found in refs [32]. The dried gels of BCZT and BSTSn were separately ground and calcined at 1000 °C for 5h. Subsequently, the pure phases of BCZT and BSTSn powders were mixed with a mole ratio of 2:3 and grinded for 1h with absolute ethanol in an agate mortar and then dried overnight. Circular green pellets with 6 mm diameter and 0.4 mm thickness were fabricated using a uniaxial hydraulic press and then sintered at 1350 °C for 7 h.

The X-ray diffraction (XRD) pattern was recorded at room temperature using the Panalytical X-Pert Pro under Cu-Kα radiation with λ ~ 1.540598 Å. The grain morphology of the sintered ceramic was observed by using the TESCAN VEGA3 Scanning Electron Microscope (SEM). The dielectric measurements were performed in the frequency range 100 Hz – 1 MHz and temperature interval from 20 °C to 250 °C, using an impedance meter HP 4284A. Polarization versus electric field (P – E) hysteresis loops were measured at 200 Hz using a ferroelectric test system (PolyK Technologies State College, PA, USA) for different temperatures.

### 3. Results and discussion

*3.1 Structural and microstructural analyses*

X-ray diffraction patterns of sintered ceramics BCZT, BSTSn, and 0.4BCZT-0.6BSTSn were recorded at room temperature (RT) within the 10–90° range, as shown in Figure 1(a). A pure perovskite structure was observed in all samples and no secondary phase was generated, indicating complete solid solubility of BSTSn within the BCZT systems. The most frequently seen diffraction peaks of barium titanate at 2θ = 45° are $(002)/(200)_T$ for the tetragonal phase, $(022)/(200)_O$ for the orthorhombic phase, and $(200)_R$ for the rhombohedral phase (fig.1b). Therefore, Rietveld refinement is an essential tool for analyzing the structural properties of all samples. The Rietveld method was used to refine the XRD patterns of the BSTSn, BCZT and 0.4BCZT-0.6BSTSn samples utilizing Fullprof software, and the results were depicted in Fig.2(a), (b), and (c). The crystal structure of the BSTSn ceramic was successfully refined in the tetragonal (T) phase of the P*4mm* space group, while the BCZT sample was refined using a mixture of tetragonal (P*4mm*) and orthorhombic (O) (A*mm2*) phases assigned to the $(022)_O$, $(200)_T$, and $(200)_O$ reflection peaks. While the mixed 0.4BCZT-0.6BSTSn solid solution was successfully refined using a combination of (T) (P*4mm*) and (O) (P*mm2*) symmetries, showing the coexistence of two structures (Tetragonal and Orthorhombic) in the matrix solid solution phase. For all ceramics, there is good agreement between observed and calculated X-ray patterns. The Rietveld refinement fitting parameters of all samples are shown in Table 1. Figure 2-d shows the evolution of the tetragonality (c/a) of all samples. It is noticeable that the c/a ratio and hence the tetragonality of the P*4mm* phase in sample BCZT-BSTSn is higher than that of the pristine BCZT sample. The increase in the tetragonality of the P*4mm* phase can be due to the lattice strain created by the incorporation of BSTSn grain into the BCZT matrix [28].

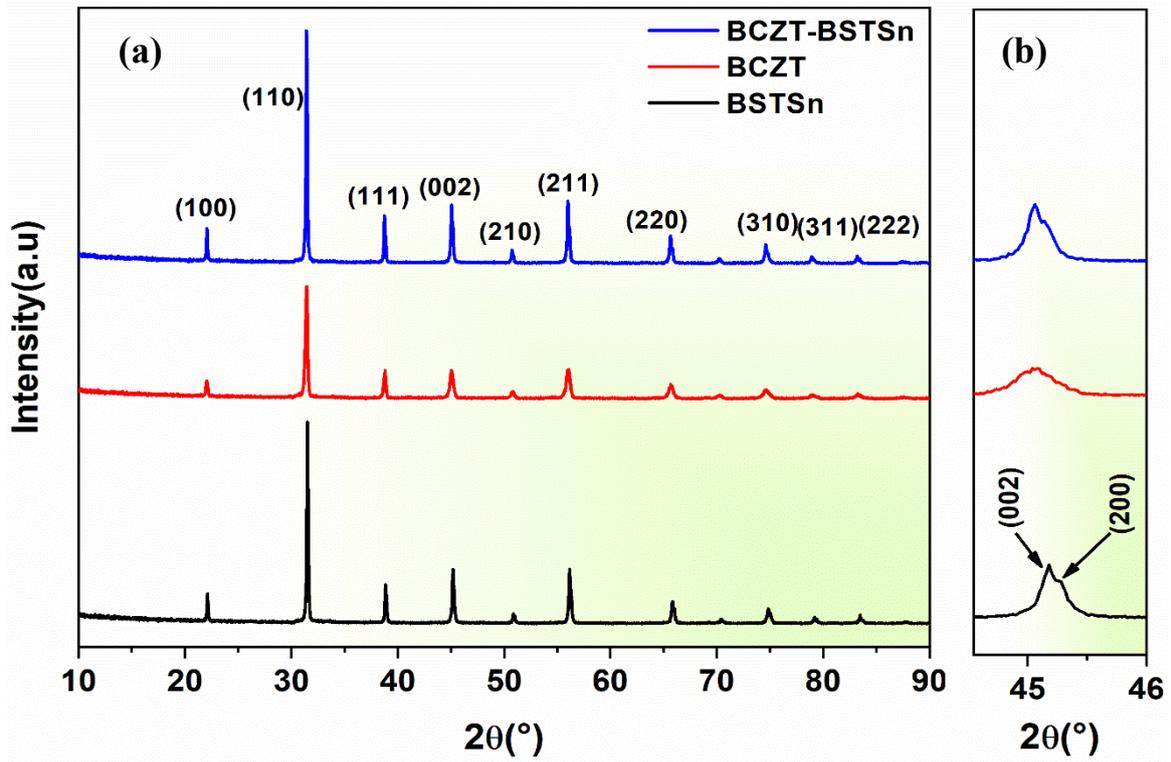

*Figure 1: (a) XRD patterns of BTSn, BCZT and 0.4BCZT–0.6BSTSn ceramics recorded at room temperature, (b) XRD patterns of enlarged (200) peaks.*

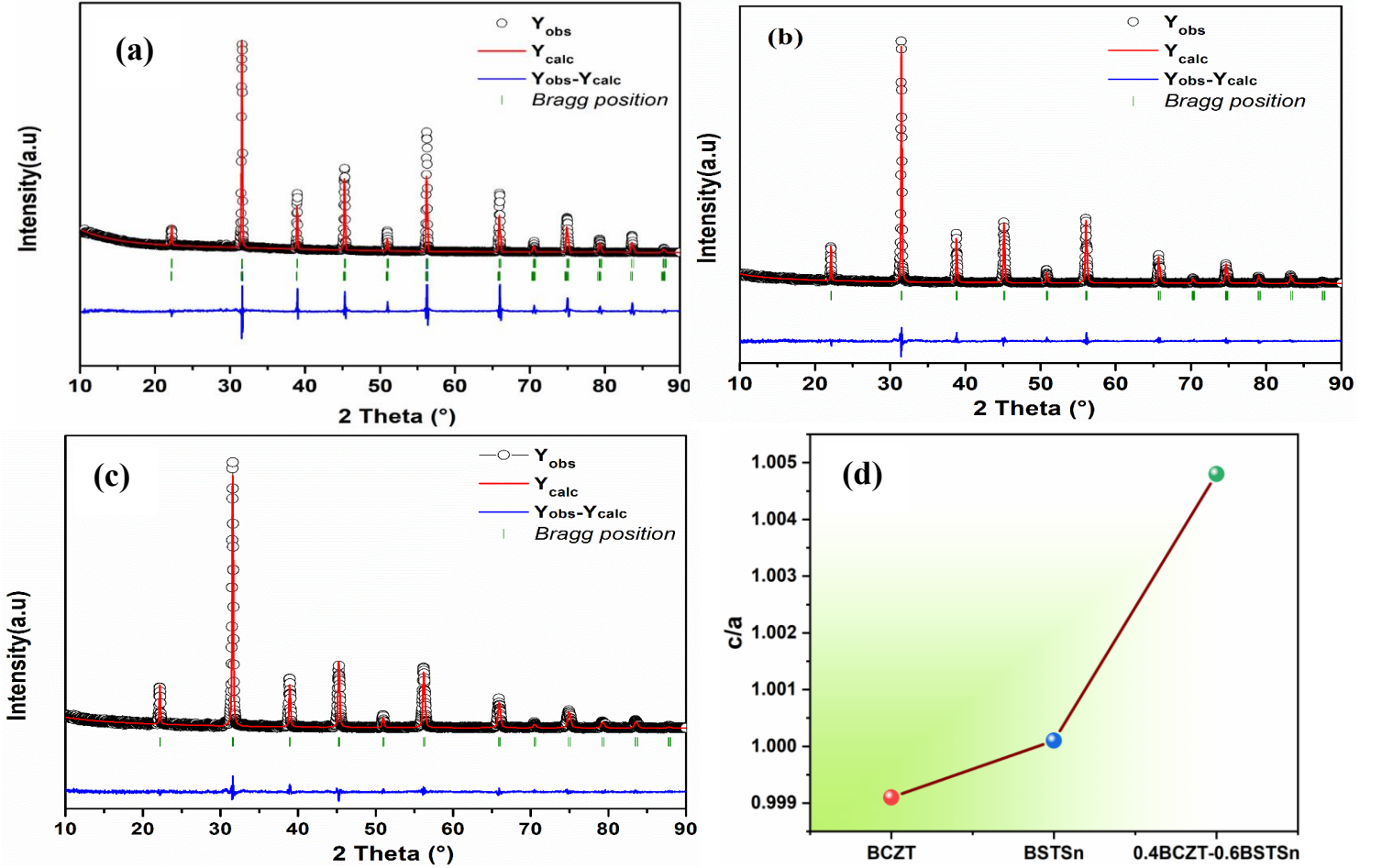

*Figure 2: Experimental X-ray diffractograms, calculated and their differences obtained for the ceramics (a) 0.4BCZT-0.6BSTSn, (b) BCZT, and (c) BSTSn. (d) c/a of all samples.*

*Table 1: Structural parameters of BSTSn, BCZT, and 0.4BCZT-0.6BSTSn obtained from Rietveld refinement.*

| Sample | Unit Cell Parameters (Å) | | | $\chi^2$ |
|---|---|---|---|---|
| | Phase 1 : *P4mm* | Phase 2 : *Pmm*2 | Phase 3 : *Amm*2 | |
| BSTSn | a = b = 4.0070<br>c = 4.0074<br>α = β = γ = 90°<br>c/a=1.00010 | - | - | 1.80 |
| BCZT | a = b = 4.0146<br>c = 4.0111<br>α = β = γ = 90°<br>c/a=0.9991 | - | a = 3.9969<br>b = 4.0143<br>c = 4.0076<br>α = β = γ = 90°<br>c/a=1.00267 | 2.78 |
| 0.4BCZT-0.6BSTSn | a = b= 3.9836<br>c = 4.0027<br>α = β = γ = 90°<br>c/a= 1.00480 | a = 4.0166<br>b = 4.0114<br>c = 4.0054<br>α = β = γ = 90°<br>c/a= 0.99721 | - | 5.46 |

Surface scanning electron microscopy (SEM) micrographs of the BSTSn, BCZT, and 0.4BCZT-0.6BSTSn ceramics, are presented in Fig. 3(a-d). These images confirm the densification of ceramics through diffusion mechanisms during the sintering process. One can see that all samples have a dense microstructure with non-uniform grain size. The average grain size was determined by image J software using the Gaussian grain distribution. For the BCZT sample, the average grain size was found to be around 17 µm which is larger than that of the BSTSn composition (7 µm). The average grain size of the 0.4BCZT-0.6BSTSn sample is about 9 µm, indicating that the incorporation of BSTSn inhibits the growth of BCZT grain. According to reports, the incorporation of Sn ions at the Ba-site and Sr in the A-site reduces the overall diffusion rate during the sintering process and may suppress the oxygen vacancies [29-31].

Fig.3e shows the variation of grain size and density of different compositions. We note that the incorporation of BSTSn in the BCZT lattice decreases the grain size and enhances the density of the 0.4BCZT-0.6BSTSn ceramic. The same result was observed in the (1 -x) $Ba_{0.85}Ca_{0.15}Zr_{0.10}Ti_{0.90}O_3$–$xBaTi_{0.89}Sn_{0.11}O_3$ ceramics synthesized by solid-state reaction [23].

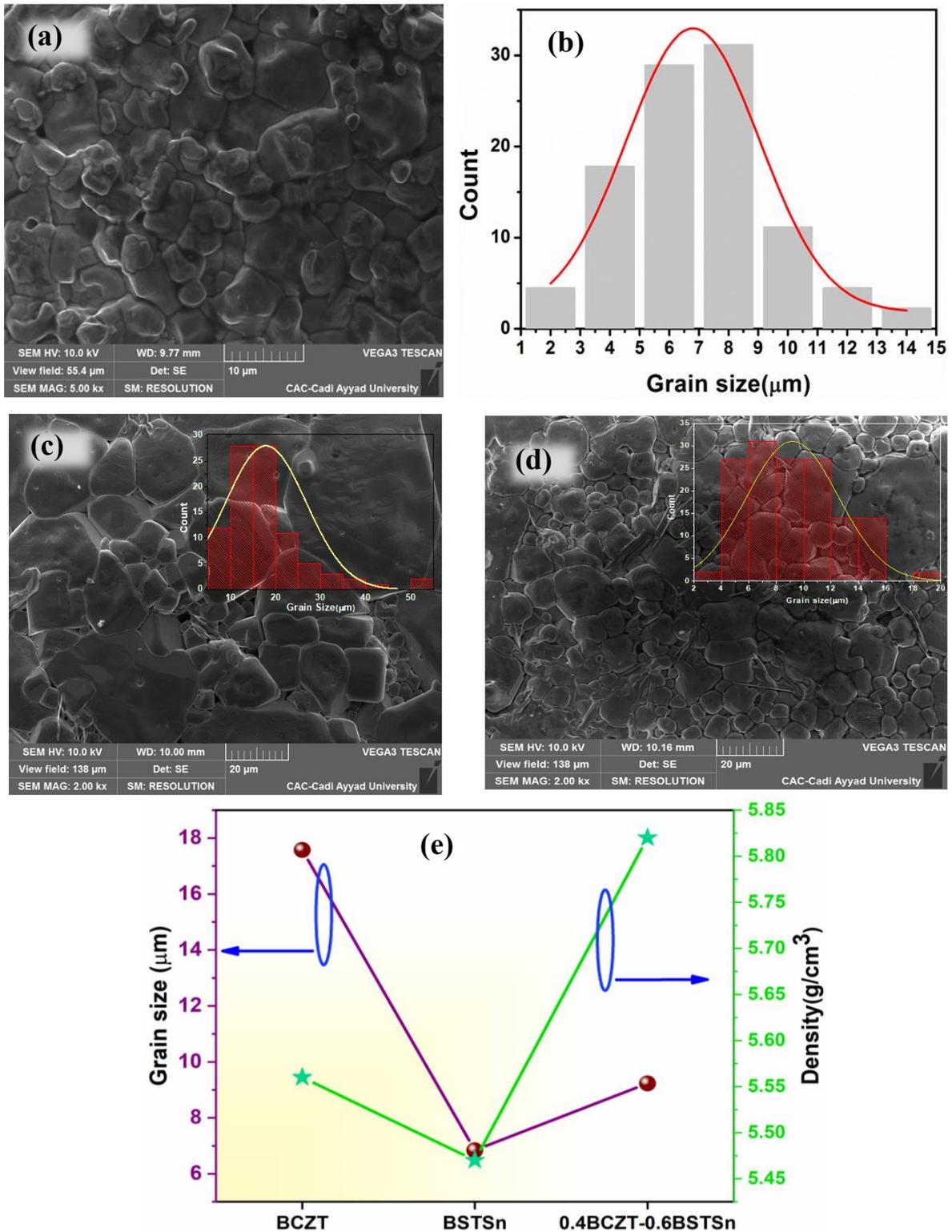

*Figure 3: SEM surface micrographs for (a,b) BSTSn, (c) BCZT and (d) 0.4BCZT-0.6BSTSn ceramics. (e) average grain size and density of all ceramics.*

*3.2 Dielectric properties*

Fig.4 displays the temperature dependence of the dielectric loss (tan δ) and the real part of the relative permittivity (ε) for all compositions measured at various frequencies. For BSTSn (Fig.

4 e,f), the dielectric constant increases with temperature and reveals a maximum value ($\varepsilon'_m$) of 5484 (for 1 kHz) at $T_C = 25$ °C, corresponding to the thermal phase transition from ferroelectric to paraelectric phase. Two distinct peaks that show conventional ferroelectric characteristics at about 32 °C and 80 °C are observed in the pure BCZT ceramic(Fig.4c,d). The two distinct transitions correspond to an Orthorhombic-Tetragonal transition (ferroelectric-ferroelectric O-T transition) around room temperature (~32°C) and a Quadratic-Cubic transition (ferroelectric-paraelectric T-C transition with a maximum of 9000) at 80°C, indicating typical ferroelectric features [20,32,33]. These results confirm the coexistence of orthorhombic and tetragonal phases in the BCZT sample in agreement with those obtained by XRD. Meanwhile, the 0.4BCZT-0.6BSTSn composition has a maximum dielectric permittivity of around 6000 at a temperature of ~50°C, corresponding to the Curie temperature (T-C transition), and relatively low dielectric losses (tgδ ≤ 0.04) compared with pure BCZT and BSTSn. Based on the literature, Sn-doped $BaTiO_3$ significantly reduces Tc and improves dielectric and energy storage properties [34-36]. Additionally, we note that the dielectric constant and dielectric loss decrease with increasing frequency, which can be attributed to the diminishing contributions of the different polarizations at high frequencies [37]. For further information on this phase transition, the Curie-Weiss law was used to fit a plot of the inverse dielectric constant as a function of temperature at a frequency of 1 kHz [38].

$$\frac{1}{\varepsilon} = \frac{(T-T_0)}{C}, \qquad (1)$$

Where $\varepsilon$ is the real part of the dielectric constant, $T_0$ and C are the Curie-Weiss temperature, and Curie-Weiss constant, respectively. In accordance with the Curie constant of the well-known displacive-type ferroelectric, such as $BaTiO_3$ ($1.7 \times 10^5$ K), the Curie constant value for all samples is around $10^5$ K [39].

In order to further characterize the dielectric relaxation behavior of BCZT, BSTSn and 0.4BCZT-0.6BSTSn ceramics, the diffusion degree (γ) is determined based on the following modified Curie-Weiss law [40]:

$$\frac{1}{\varepsilon} - \frac{1}{\varepsilon_m} = \frac{(T-T_m)^\gamma}{C} \quad (1 < \gamma < 2), \qquad (2)$$

Where ε, $\varepsilon_m$ are the real part of the dielectric constant and its maximum value, respectively, and γ is the degree of diffusion of the transition. Eq (2) indicates an ideal relaxor and depicts the so-called full diffuse phase transition (DPT) for γ = 2, whereas for γ = 1, it fits a typical ferroelectric. In contrast, an "incomplete" DPT is characterized by intermediate values of γ between 1 and 2 [41].

The γ value of the o.4BCZT-0.6BSTSn sample is higher than γ of pure BSTSn and lately lower than γ of pristine BCZT. The slight decrease in the γ of 0.4BCZT-0.6BSTSn sample value is likely due to the effect of the B-site sublattice and a large number of $Sr^{2+}$ vacancies created in this sample [31,42]. The dielectric properties of all studied ceramics are summarized in Table 2.

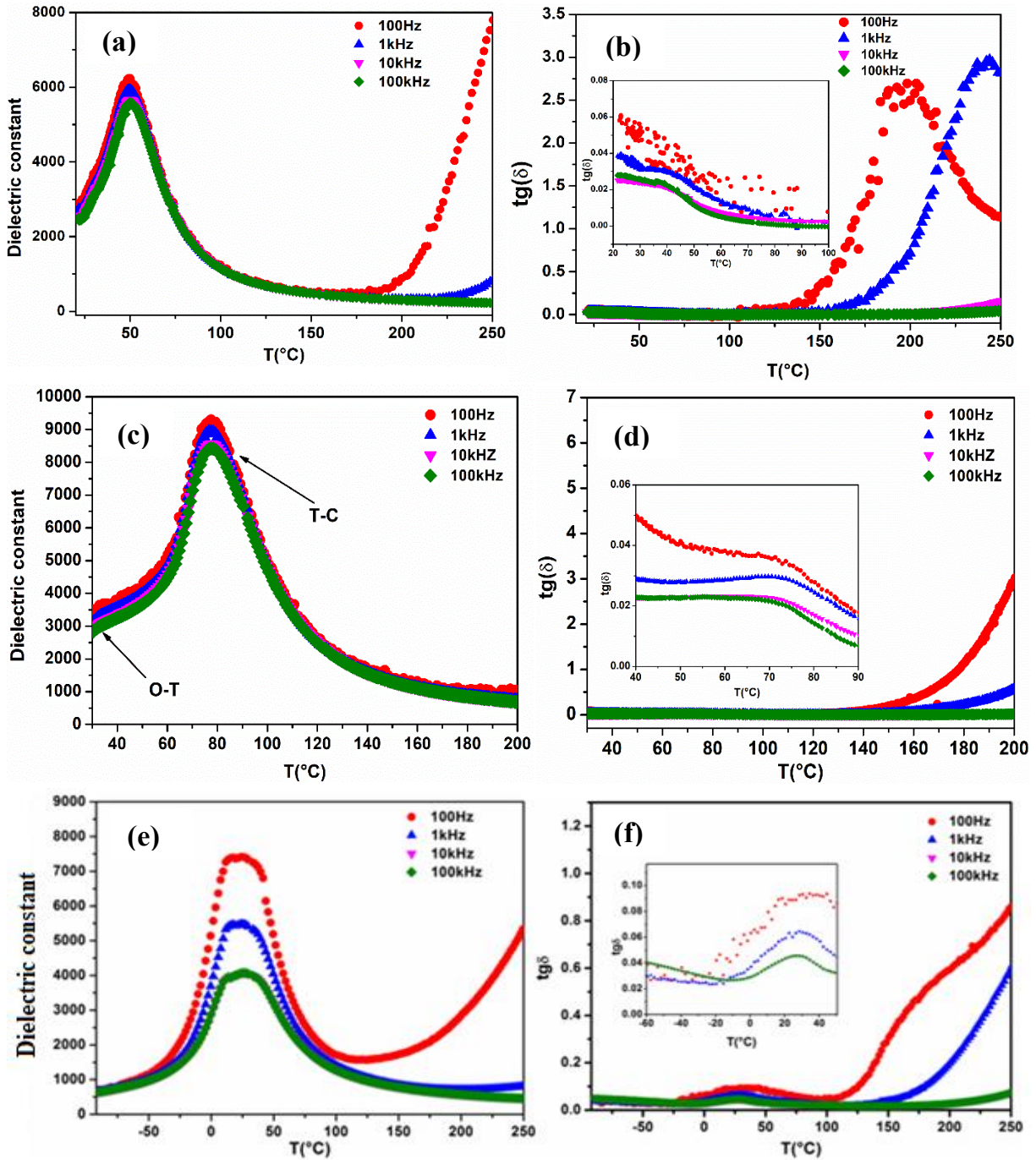

*Figure 4: The temperature-dependent dielectric performance of the (a,b) 0.4BCZT-0.6BSTSn, (c,d) BCZT and (e,f) BSTSn ceramics.*

*Table 2: Dielectric properties of BCZT, BSTSn and 0.4BCZT-0.6BSTSn ceramics at 1 kHz.*

| Material | $\varepsilon_m$ | $T_0(°C)$ | $T_m(°C)$ | $C\times10^5$ | $T_{dev}$ | $\Delta T_m$ | $\gamma$ |
|---|---|---|---|---|---|---|---|
| BCZT | 9000 | 75 | 78 | 1.07 | 105 | 30 | 1.73 |
| BSTSn | 5484 | 26 | 23.6 | 1.04 | 52.81 | 29.21 | 1.62 |
| 0.4BCZT-0.6BSTSn | 6000 | 47.67 | 50 | 1.09 | 77.8 | 30.83 | 1.71 |

*3.3 Ferroelectric and energy storage properties*

To study the ferroelectric characteristics of all investigated compositions (BCZT, BSTSn, and 0.4BCZT–0.6BSTSn), the thermal evolution of P-E hysteresis loops close to the Curie temperature was measured using a maximum electric field of 30 kV/cm at 200 Hz. Fig. 4(a)-(c) exhibit the bipolar P-E loops of BCZT, BSTSn, and 0.4BCZT–0.6BSTSn ceramics. It has been observed that all samples demonstrate typical ferroelectric P-E hysteresis loops. Indeed, each composition has undergone a FE-PE phase transition, as indicated by the hysteresis curve becoming thinner and changing to a linear response as temperature rises and paraelectric domains are created, which is consistent with dielectric results. As exhibited in the bipolar hysteresis loops, the maximum polarization $P_{max}$ increases significantly twice in 0.4BCZT-0.6BSTSn in comparison to BCZT and BSTSn pure (Fig 5-(d)). The enhancement of $P_{max}$ in the 0.4BCZT-0.6BSTSn sample could be due to the reduction in grain size, which is directly related to the lower interfacial polarization [43]. On the other hand, the coercive filed $E_c$ values are slightly decreased as compared with pure BCZT, which indicates that the ferroelectric domains can switch polarizations more easily [44].

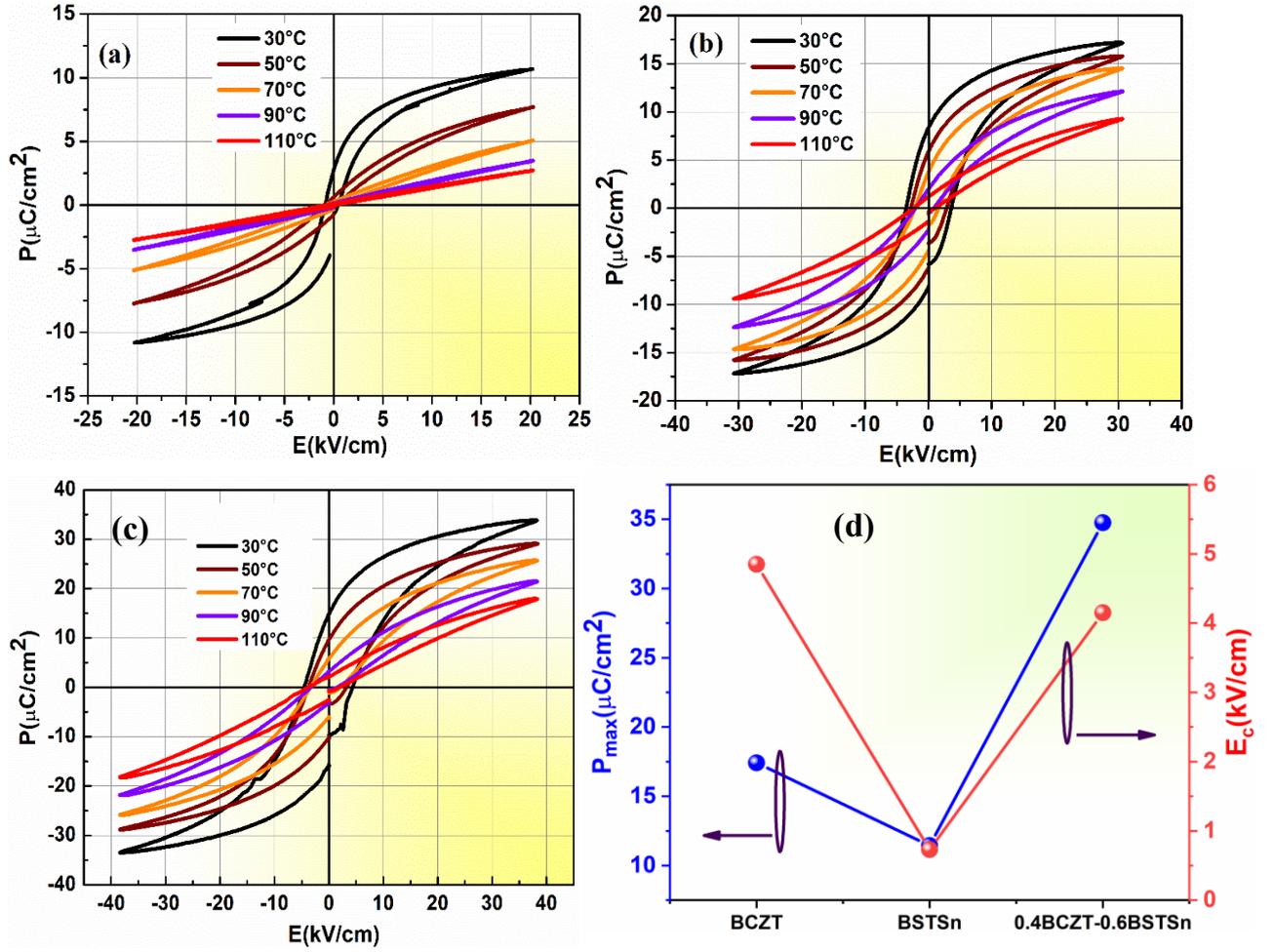

*Figure 5: P-E hysteresis loops at different temperatures of (a) BSTSn, (b) BCZT and (c) 0.4BCZT-0.6BSTSn. (d) $P_{max}$ and $E_c$ values of all samples at room temperature.*

It is well-known that energy storage performances such as total energy density ($W_{tot}$), recoverable energy density ($W_{rec}$), and energy storage efficiency ($\eta$), of dielectric materials, can be extracted from their P–E loops (Fig.5(a)-(c)). The energy storage parameters were calculated using the following equations [45]:

$$W_{tot} = \int_0^{Pmax} E dP, \tag{3}$$

$$W_{rec} = \int_{Pr}^{Pmax} E dP, \tag{4}$$

$$\tag{5}$$

$$\eta(\%) = \frac{W_{rec}}{W_{tot}} * 100 = \frac{W_{rec}}{W_{rec}+W_{loss}} * 100,$$

The various energy storage parameters of BCZT, BSTSn, and 0.4BCZT-0.6BSTSn ceramics have been calculated and presented in Fig.6 (a)-(c). Furthermore, the maximum values calculated for $W_{rec}$ are found to be around 99.9 mJ/cm$^3$ at 101°C, 58.8 mJ/cm$^3$ at 45°C, and 255.4 mJ/cm$^3$ at 90°C for BCZT, BSTSn and 0.4BCZT-0.6BSTSn respectively. Under an applied electric field of 30 kV/cm, the BSTSn sample exhibits a higher $\eta$ of 84.4% than those of BCZT (74.1%) and 0.4BCZT-0.6BSTSn (66.9%). Notably, higher efficiency results in less energy loss during the charging and discharging processes. According to the calculation above (Eq. 4), the more important the difference between P$_{max}$ and P$_r$, the better the energy storage performances. One can see that the 0.4BCZT-0.6BSTSn sample shows the highest $W_{rec}$ among all other samples (Fig.6(e)). As reported in previous research, the coexistence of multiple stages enhances the dielectric properties and thus improves the energy storage density [28,46,47]. These results indicate that the used medium entropy method is an effective way to improve the energy storage density through decreasing the grain size and creating of multiphases, which leads to break the long range ferroelectrics (macro domains) and generating the PNRs (micro domains) [10].

For ceramic capacitor applications, temperature stability is a crucial evaluation characteristic since energy storage devices frequently operate in harsh and/or hot environments. In addition, the thermal stability of energy storage performance is critical for the practical application of electrical devices [29,48,49]. The dependence of thermal variation of the energy storage can be determined using the following equation:

$$\frac{\Delta W_{rec,T}}{W_{rec,300K}} = \left|\frac{W_{rec,T}-W_{rec,300K}}{W_{rec,300K}}\right| \qquad (6)$$

Where, $W_{rec,T}$ is the $W_{rec}$ value at a given temperature, and $\Delta W_{rec,T}$ is the difference between $W_{rec,T}$ and $W_{rec,300\ K}$. Fig.6 (d) shows the thermal evolution of the recovered energy storage density of the sample 0.4BCZT-0.6BSTSn. The 0.4BCZT-0.6BSTSn sample exhibits good thermal energy storage stability of less than 10% ($W_{rec}$ ~ 205.6-225.4 mJ/cm$^3$) in the temperature range of 40 - 120°C.

To set out our results to the literature, Table 3 summarizes the comparison of the recoverable energy density and the energy storage efficiency of various published lead-free ceramics. It is well known that energy storage performance can be affected by elaboration methods, chemical composition, grain size engineering, and applied electric fields. The study's findings suggest that medium-entropy engineering is an effective approach to improve energy storage capacity for designing novel high-performance ceramic capacitors.

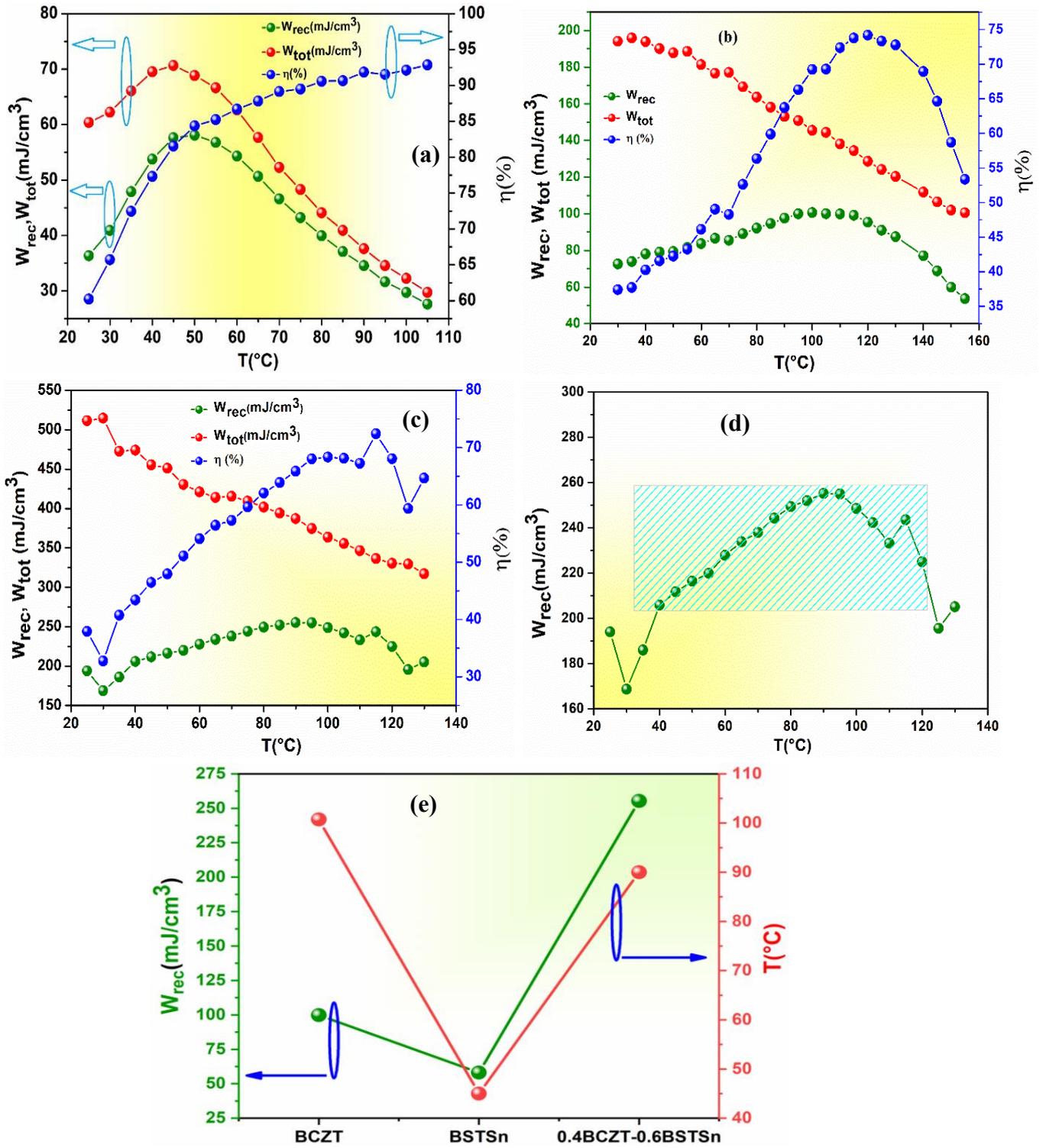

*Figure 6: (a)-(c) Thermal evolution of energy storage performances of BCZT, BSTSn and 0.4BCZT-0.6BSTSn ceramics. (d) Temperature dependence of $W_{rec}$ of the 0.4BCZT-0.6BSTSn sample. (e) The maximum of $W_{rec}$ and temperature values of different samples.*

*Table 3: Comparison of energy storage performances of 0.4BCZT–0.6BSTSn ceramic with other lead-free ferroelectric ceramics.*

| Matériau | $W_{rec}$ (mJ/cm$^3$) | E (kV/cm) | $\eta$ (%) | T (°C) | Refs. |
|---|---|---|---|---|---|
| **BCZT** | **99.9** | **30** | **74.1** | **100.7** | **This work** |
| **BSTSn** | **58.1** | **20** | **84.4** | **45** | **This work** |
| **0.4BCZT-0.6BSTSn** | **255.4** | **38** | **66.9** | **90** | **This work** |
| $Ba_{0.85}Ca_{0.15}Zr_{0.10}Ti_{0.90}O_3$ | 62 | 25 | 72.9 | 130 | [20] |
| $Ba_{0.85}Ca_{0.15}Zr_{0.10}Ti_{0.90}O_3$ | 14 | 6.5 | 80 | 129 | [50] |
| $Ba_{0.98}Ca_{0.02}Zr_{0.02}Ti_{0.98}O_3$ | 82 | 20 | 51.5 | - | [51] |
| $Ba_{0.85}Ca_{0.15}Zr_{0.10}Ti_{0.90}O_3$ | 128.8 | 40 | 55.4 | RT | [52] |
| $Ba_{0.85}Ca_{0.15}Zr_{0.10}Ti_{0.90}O_3$ | 121.6 | 60 | 51.7 | RT | [53] |
| $BaZr_{0.05}Ti_{0.95}O_3$ | 218 | 50 | 72 | RT | [54] |
| $BaTi_{0.89}Sn_{0.11}O_3$ | 85.1 | 72.4 | 25 | 85.07 | [34] |
| $BaTi_{0.89}Sn_{0.11}O_3$ | 71.3 | 25 | 67.9 | 28 | [55] |
| 0.4BCZT–0.6BTSn | 137.86 | 30 | 86.19 | 80 | [25] |
| $0.85[(1-x) Bi_{0.5}Na_{0.5}TiO_3–xBaTiO_3]–0.15Na_{0.73}Bi_{0.09}NbO_3$ | 1100 | 122 | 67.9 | - | [56] |

## 3.4 Electrocaloric effect properties

The electrocaloric (EC) behavior of a material can be characterized by its adiabatic temperature change (ΔT), and isothermal entropy change (ΔS). These parameters can be determined through various methods: direct measurements of ΔT, quasi-direct measurements of heat, which typically provide ΔS (entropy change), and indirect methods, where ΔS is derived from isothermal measurements of the electrical polarization (P) as a function of the electric field (E), or ΔT is obtained from adiabatic polarization measurements. In the present work, the reversible adiabatic temperature change (ΔT) and the isothermal entropy change (ΔS) were deduced via indirect method [57].

The electrocaloric effect (ECE) in all ceramics is evaluated using the indirect Maxwell method [58]. The EC effect is calculated based on the measured ferroelectric order parameter P(E, T), obtained from the upper branches of the corresponding P–E hysteresis loops measured at 200 Hz. The adiabatic temperature change (ΔT) and the isothermal entropy change (ΔS) induced by varying the applied electric field from $E_1=0$ to $E_2$ in the EC material are expressed as follows:

$$\Delta T = -\frac{1}{\rho} \int_{E_1}^{E_2} \frac{T}{C_p} \left(\frac{\partial P}{\partial T}\right)_E dE \qquad (7)$$

$$\Delta S = -\frac{1}{\rho} \int_{E_1}^{E_2} \left(\frac{\partial P}{\partial T}\right)_E dE \qquad (8)$$

Where ρ is the density of the ferroelectric sample which was measured by the Archimed method and the value of 5.82 g/cm³, $C_p$ is the heat capacity obtained by integrating the heat flow curves in the measured range. $E_1$ and $E_2$ are the starting and final applied fields, respectively. P is the polarization. The critical factor $(\partial P/\partial T)_E$ is calculated by applying a seventh-order polynomial fit to the raw P–T data obtained under various external electric fields from the hysteresis loops. The electrocaloric temperature change (ΔT) and the isothermal entropy change (ΔS) for 0.4BCZT-0.6BSTSn are then determined using the equations (7) and (8). The corresponding results are presented in Fig.8-(a) and (b).

For all samples, ΔT rises with increasing E, and its maxima shift slightly to higher temperatures. The BCZT and BSTSn pristine ceramics show ΔT values of 0.54 K and 0.63 K under electric fields of 30 and 20 kV/cm, respectively. On the other hand, the sample 0.4BCZT-0.6BSTSn demonstrates large values of ΔT = 1.36 K under E of 30 kV/cm. All samples' thermal variation of ΔT peaks around the ferroelectric-paraelectric phase transition. The large EC temperature change in this material composite is attributed to the multiphase coexistence. This phenomenon is thought to arise from the higher number of polar states present in the multiphase structure compared to other compositions [59]. Notably, a broad ΔT peak was observed for 0.4BCZT-0.6BSTSn, driven by its diffused phase transition and demonstrating excellent electrocaloric (EC) strength [60]. Furthermore, near the critical point, the electric field reduces the energy barrier for polarization rotation, resulting in a large entropy change [59]. Besides, as the electric field increases, the EC temperature change (ΔT) becomes more pronounced, and the EC peak shifts slightly toward higher temperatures, which is in agreement with the variation of the polarization versus temperature curves with the electric field [61].

Additionally, electrocaloric responsivity ($\xi_{max}$) indicates a material's capability to alter its EC temperature in response to an applied electric field and is defined as $\xi_{max} = (\Delta T_{max}/\Delta E_{max})$[62]. For industrial applications, achieving high electrocaloric (EC) efficiency is essential when exploring suitable lead-free electrocaloric materials [63]. Under a low electric field of 30 kV/cm, it was observed that the 0.4BCZT-0.6BSTSn exhibited a high EC efficiency of 0.453 Kmm/kV.

In comparison to other materials, our work demonstrates the highest electrocaloric responsivity ($\xi_{max}$) of 0.453 K·mm/kV using the indirect method. This indicates a more efficient electrocaloric effect compared to other materials such as $BCS_xT$ (x = 0.20) with 0.16 K·mm/kV [64], $Ba_{0.8}Ca_{0.2}Zr_{0.04}Ti_{0.96}O_3$ with 0.34 K·mm/kV [65], $0.7BaZr_{0.2}Ti_{0.8}O_3$-$0.3Ba_{0.7}Ca_{0.3}TiO_3$ with 0.15 K·mm/kV [66], $BaTi_{0.895}Sn_{0.105}O_3$ with 0.31 K·mm/kV [67], $BaZr_{0.2}Ti_{0.8}O_3$ (ceramic + glass) with 0.31 K·mm/kV (direct method) [68], and $PbMg_{1/3}Nb_{2/3}O_3$ with 0.27 K·mm/kV (direct method) [69]. This highlights the potential of our material for industrial applications requiring high electrocaloric efficiency.

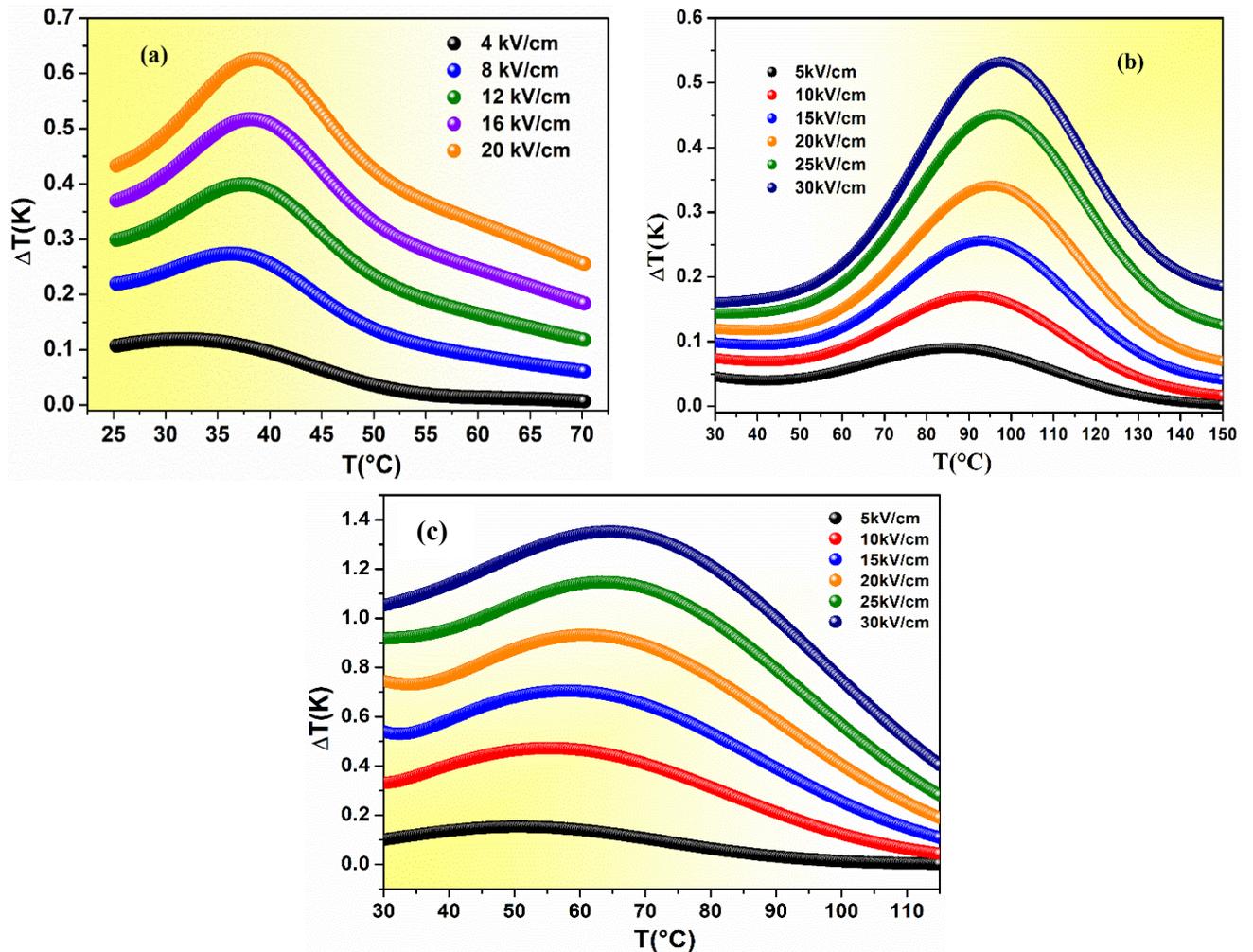

*Figure 8: Temperature dependence ΔT of (a) BSTSn, (b) BCZT, and (c) 0.4BCZT-0.6BSTSn ceramics at different applied electric fields.*

## 4. Conclusion

In summary, the BCZT, BSTSn, and 0.4BCZT-0.6BSTSn lead-free ferroelectric ceramics were successfully prepared through a sol-gel method, and BSTSn was introduced into BCZT ceramics to construct medium-entropy material. The investigation of energy storage performances of 0.4BCZT-0.6BSTSn ceramic demonstrated the improved values of $W_{rec}$ of 255.38 mJ/cm$^3$ and $\eta$ of 66.91% under a low electric field. In addition, excellent temperature stability (40–120 °C) of $W_{rec}$ (less than 10%) was achieved in the 0.4BCZT-0.6BSTSn sample. Furthermore, a large electrocaloric effect temperature change of $\Delta T = 1.36$ K under a low electric field of 30 kV/cm was extracted from the 0.4BCZT-0.6BSTSn sample. All the above results suggest that the 0.4BCZT-0.6BSTSn ceramic should be an environmentally friendly candidate for refrigeration and energy storage applications around the RT.

**Acknowledgements**

This work was supported by the H-GREEN project (No 101130520) and the European Union's Horizon – MSCA-SE-2021 Research and Innovation Programme under the Marie Sklodowvska-Curie EWALKD_Grant Agreement 101086250. The Slovenian Research and Innovation Agency (research core funding P2-0105 and research project N2-0212) is acknowledged.